\documentclass[12pt,pra,article,showpacs,array,amscd,amsmath,hhline]{revtex4}
\usepackage{graphicx}
\usepackage{graphics}
\begin{document} 


\newcommand{\eq}{\begin{equation}}
\newcommand{\eqe}{\end{equation}}

\title{Analytic solutions for the one-dimensional compressible Euler equation with heat conduction  
closed with different kind of equation of states}

\author{Imre F. Barna$^a$ and L\'aszl\'o M\'aty\'as$^b$}
\address{$^a$ Energy Research Centre of the Hungarian Academy of Sciences \\ 
H-1525 Budapest, P.P. Box 49, Hungary \\ 
$^b$  Sapientia University, Department Bioengineering, 
Libert\u{a}tii sq. 1, 530104 Miercurea Ciuc, Romania}

\date{\today}
\begin{abstract}   We present analytic self-similar or traveling wave solutions for a one-dimensional 
coupled system of continuity, compressible Euler and heat conduction equations. Different kind of equation of states  
are investigated. 
In certain forms of the equation of state one can arrive to a picture 
regarding the long time behavior of density and pressure. The impact of these quantities on the evolution of temperature is also discussed. 
\end{abstract}

\pacs{34.10.+x, 34.50.-s, 34.50.Fa}

\maketitle

\section{Introduction}
Studying hydrodynamical systems with heat conduction is an important 
task, both from application and from theoretical side as well. 
There are numerous monographs and textbooks available which investigate 
three dimensional real fluid mechanics with heat transfer from the engineering point of view \cite{tannes}.
Usually, all the technical questions like the applied numerics, grids are analyzed in depth as well. 

In the following we will investigate the most simple one dimensional system of continuity, compressible and non-viscous 
Euler and heat conduction equations. We try to find analytic results, as a first point self-similar physically important diffusive solutions are investigated.  If such solutions are not to find, as a second point  traveling wave solutions  are obtained and discussed. 

These are basically two distinct very powerful trial functions to investigate the global properties of the solutions of various  single non-linear partial differential equations  \cite{sedov,barenb,zeld, kers}.  One is the so-called 
self-similar solution which describes the intermediate asymptotics of a problem: they hold when the precise initial 
conditions are no longer important, but before the system has reached its final steady state. They are much simpler than 
the full solutions and so easier to understand and study in different regions of parameter space. A final reason for studying them is that they are solutions of a system of ordinary differential equations and hence do not suffer the extra inherent numerical problems of the full partial differential equations. In some cases self-similar solutions helps to understand diffusion-like properties or the existence of compact supports of the solution. 

The other method is the traveling wave Ansatz which gives us a deeper insight into the wave properties of the system like 
the propagation speed which can be even time dependent. 

In our former work the paradox of heat conduction was investigated with the self-similar Ansatz  \cite{barna} 
and gave a new equation which has a finite propagation speed. 
Recently, a study about the generalized Cattaneo law is under publication where both the 
heat conduction coefficient and relaxation time have power law temperature dependence \cite{barna1}. 
The results can be continuous or shock wave-like with compact support.   
With a straight forward generalization of the self-similar Ansatz even a partial differential equation(PDE) system  the 3 dimensional Navier-Stokes equations was successfully investigated \cite{barna2}. 
Gilding and Kersner \cite{kers}  studied large number of nonlinear diffusion-convection problems with the help of traveling waves.  

As a beginning, without completeness we mention some analytic studies about hydrodynamical systems which are 
available in the literature.

Beyond  traveling waves and self-similar Ans\"atze there are various examination techniques available in the literature.  
Manwai  \cite{manwai} studied the N-dimensional $(N \ge 1)$ radial 
Navier-Stokes(NS) and Euler equation with different kind of viscosity and pressure 
dependences and presented analytical blow up solutions.	
His works are still 1+1 dimensional (one spatial and one time dimension)
investigations.  Another well established and popular investigation method is based on 
Lie algebra by which numerous studies are available.  
Some of them are even for the three dimensional case,  
for more see \cite{lie}. Unfortunately, no explicit solutions are shown and 
analyzed there. Fushchich {\it{et al.}} \cite{fus} construct a complete set of ${\tilde{G}}(1,3)$-inequivalent Ans\"atze of codimension 1 for the NS system and they present 19 different analytical solutions for one or two space dimensions. 
They last solution is very closed to our former one \cite{barna2}  but not identical. 
Further two and three dimensional studies of the Navier-Stokes equation based on group analytical method were presented by Grassi \cite{grassi}. They also presented solutions which look almost the same as ours \cite{barna2}, but they consider only two space dimensions.  

 Recently, Hu {\it{et al.}} \cite{hu} presents a study where 
symmetry reductions and exact solutions of the (2+1)-dimensional NS were presented. 
Aristov and Polyanin \cite{arist} use various methods  like Crocco transformation,  generalized separation of variables or the method of functional separation of variables for the NS and present large number of new classes of exact solutions. 
Sedov in his classical work \cite{sedov} (Page 120) presented analytic solutions for the three dimensional spherical NS equation 
where all three velocity components and the pressure have polar angle dependence ($\theta$) only. Even this kind of restricted symmetry led to a non-linear coupled ordinary differential equation system which has a very rich mathematical structure. Some similarity reduction solutions of the two dimensional incompressible NS equation was presented by 
Xia-Ju \cite{jiao}.  Additional solutions are available for the 2+1 dimensional NS also via symmetry reduction techniques by 
\cite{fakhar}.  
There is a full three dimensional Lie group analysis is available for the three dimensional Euler equation of gas dynamics, 
with polytropic EOS unfortunately without any heat conduction mechanism \cite{irani}. 
Of course one may find numerical methods for solving equations of fluid dynamics \cite{ZiTaNi05}.
 The topic is quite important, there are interesting  applications of the equations of fluids in granular matter \cite{GrSaViPu11}.     
 
To our knowledge there are very few analytic (self-similar or traveling wave) solutions known for any non-linear partial 
differential equation systems (PDEs) till today \cite{robi}.  Our experience shows that there are no such comparative studies available  for fluid mechanics with heat conduction in the literature  which we present on the next pages.  

In the next section we outline our starting model in the later subsections we give various solutions for three different polytropic equation of states (EOS),  plus for the virial EOS. If our results meet some other theories (like porous media equation) 
then the corresponding literature is mentioned.     
The paper ends with a short summary. 
 
\section{The basic model} 
We start with the following one-dimensional compressible and non-viscous fluid with ordinary heat conduction  which means  
the following coupled PDE system of continuity, Euler and heat conduction
\begin{eqnarray}
\rho(x,t)_t + [\rho(x,t)v(x,t)]_x =0, \nonumber \\
v(x,t)_t + v(x,t)v(x,t)_x = -\frac{1}{\rho(x,t)} p(x,t)_x, \nonumber \\
T(x,t)_t + v(x,t)T(x,t)_x = \lambda T(x,t)_{xx},
\label{pde}
\end{eqnarray}
where $\rho, v, T, P$ are the density, velocity, temperature and pressure 
field or distribution, respectively. Subscript means partial derivation 
with respect of time and the $x$ to the Cartesian coordinate. 
We skip viscous velocity term in Euler and in the energy equation as well. 
As a starting point we include the polytropic EOS $p(x,t) = a \rho(x,t)^n$ were $n$ is a real number and $a$ is a material constant and $\lambda$ is the heat conduction coefficient. 
After the investigation of this basic equation system we will introduce modifications for the equation of state. 

In the following we are looking for the solution in the form of the
self-similar Ansatz which is well-known from \cite{sedov,barenb,zeld} 
\eq 
V(x,t)=t^{-\alpha}f\left(\frac{x}{t^\beta}\right):=t^{-\alpha}f(\eta) 
\label{self}
\eqe 
where $V(x,t)$ can be an arbitrary variable of a PDE and $t$ means time and $x$ means spatial 
dependence.
The similarity exponents $\alpha$ and $\beta$ are of primary physical importance since $\alpha$  represents the rate of decay of the magnitude $V(x,t)$, while $\beta$  is the rate of spread 
(or contraction if  $\beta<0$ ) of the space distribution as time goes on.
The validity of this self-similar Ansatz also means that the solution has no 
characteristic time scale.  The most powerful result of this Ansatz is the fundamental or 
Gaussian solution of the Fourier heat conduction equation (or for Fick's
diffusion equation) with $\alpha =\beta = 1/2$.  The function $f(\eta)$ is called the shape function. 

For our system we use the following notations and shape functions
\eq
T(x,t) = t^{-\alpha}f\left(\frac{x}{t^{\beta}} \right), \>\>
v(x,t) = t^{-\delta}g\left(\frac{x}{t^{\beta}} \right), \>\> 
\rho(x,t) =  t^{-\gamma}h\left(\frac{x}{t^{\beta}} \right)
\label{ans}
\eqe
where the new variable is $\eta = x/t^{\beta}$. 
Calculating all the first time and spatial derivatives of (\ref{ans}) and putting them back to (\ref{pde}) after some algebra we get the following 
non-linear ordinary differential equation (ODE) system
\begin{eqnarray}  
-\gamma h - \frac{1}{2} \eta h' + g h' + hg' = 0, \nonumber \\
-\frac{1}{2} g - \frac{1}{2} \eta g' + gg' = -a 
\left(\frac{1+\gamma}{\gamma} \right) h^{\left(\frac{1+\gamma}{\gamma}-2 \right) } h', \nonumber \\ 
-\alpha f - \frac{1}{2}\eta f' +gf' = \lambda f'', 
\end{eqnarray}
where prime means derivation with respect to $\eta$. 
Among the initially free parameters $\alpha,\beta,\gamma, n$ we got 
the following constraints $\alpha, \gamma$ are still arbitrary 
but $\beta = \delta = 1/2 $ and $n = (1+\gamma)/\gamma$. 
The material constants $c,\lambda$ are still independent.  
Note, that the first and second equations of (\ref{pde}) are independent of
the third one. 

\subsection{Aspects coming from the conservation law of density}
The first equation is a conservation equation (without source term) and if $\gamma =1/2$ it can be integrated 
\eq
 \left(\frac{\eta h}{2} \right)' = (gh)' 
\eqe 
giving $g = \eta/2$ and $g' = 1/2$ when $h \ne 0 $ and the integration 
constant was chosen zero. 
Plugging this result back into the second equation of (\ref{pde}).  
\eq
-\frac{\eta}{4} = -3ahh' 
\label{2}
\eqe
where $n=(1+\gamma)/\gamma = 3$. 
Which means that only the $p = a\rho^3$ ploytropic EOS can give self-similar 
solutions in this flow system. 
The solution of (\ref{2}) can be obtained via a simple integration giving the following expression 
\eq
h = \sqrt{ \left( \frac{4\eta^2 + 2c_1}{3a}  \right)}
\eqe
where $c_1$ is the integration constant (we will set it to 0) 
and $a$ is still the material constant of the polytropic EOS. 
Last, let's investigate the third heat conduction equation, with 
the knowledge of the velocity its solution becomes quite simple 
\eq
-\alpha f = \lambda f''.
\eqe
The solution is well known 
\eq
f = c_2 cos \left( \frac{\alpha \eta}{\lambda} \right) + 
c_3 sin \left( \frac{\alpha \eta}{\lambda} \right)  
\eqe
At last, we summarize all our results 
\eq
g = \frac{\eta}{2}, \>\> h = \sqrt{ \left( \frac{4\eta^2 + 2c_1}{3a}  \right)}, \>\> f = c_2 cos \left( \frac{\alpha \eta}{\lambda} \right) + 
c_3 sin \left( \frac{\alpha \eta}{\lambda} \right)
\eqe
where $n =3$, $\gamma = \beta = \delta = 1/2$ and $\alpha$ is arbitrary. 
Now the velocity, density and the temperature field read 
\begin{eqnarray}
v(x,t) = t^{-\delta}g(\eta) = \frac{x}{2t}, \nonumber \\ 
\rho(x,t) = t^{-\gamma}h(\eta) = t^{-\frac{1}{2}} 
\sqrt{  \frac{ 4 \left (\frac{x}{\sqrt{t}} \right)^2 + 2c_1}{3c}  }, \nonumber  \\
T(x,t) = t^{-\alpha}g(\eta) = t^{-\alpha} \left [c_2 cos \left( \frac{\alpha x }{\lambda \sqrt{t}} \right) + 
c_3 sin \left( \frac{\alpha x}{\lambda \sqrt{t} } \right) \right]
\end{eqnarray}
and the applied polytropic EOS  
\eq
p = a\rho^n = a (t^{-\gamma}h(\eta))^3 = a \left[ t^{-\frac{1}{2}}  
\sqrt{  \frac{ 4 \left (\frac{x}{\sqrt{t}} \right)^2 + 2c_1}{3c}  } \right]^3. 
\eqe
 All the solutions have physical meaning for correspondingly bounded space 
and sufficiently large times.  
We will see later on that according to the continuity equation 
the velocity field is always a simple function of time and coordinate, 
however the temperature distribution have a much complexer form.  In this case  it is  a quickly oscillating and decaying function or a slowly oscillating and slowly decaying function.
   
\subsection{Aspects on solution starting from 
the equation regarding the velocity}

We had the possibility to see that in certain cases, when the first equation 
is integrable,  solutions of the system can be found. This was accessible when 
the pressure is proportional to the third power of the density. 
Now we try to find solutions when the pressure is proportional to the second
power of the density
\eq
p=\frac{b}{2}\rho^2.
\eqe  
The role of $b$ is to fix the proper physical dimension. 
In this case the second equation of  (\ref{pde}) can be written 
\eq 
v_t + vv_x = -  b  \rho_x
\eqe
If we consider that the system is not too far from the steady state, 
correspondingly the change in time of the gradient of the density 
is negligible, we get
\eq
v=- b\rho_x \cdot t + \hat{c}, 
\label{vtransient}
\eqe
where $\hat{c}$ can be related to initial velocity and initial gradients at time 
$t_0$ if these terms do not cancel each other.  
We suppose that (\ref{vtransient}) holds for a longer time scale than 
the gradient would undergo to considerable changes. We are interested in 
a density profile in such a quasi-stationary state.  
The procedure can be also regarded as taking a first initial condition of 
the velocity field and iterate it successively \cite{Micula92}. 
Inserting  (\ref{vtransient})  
into the mass conservation formula we get
\eq 
\rho_t 
-t b (\rho_x)^2 
+ \hat{c}\rho_x -t b  \rho  \rho_{xx}  
=0.
\eqe 
If $\hat{c}=0$ and dividing the equation with $t$ we obtain the following
equation 
\eq
\frac{\rho_t}{ t} 
-  b (\rho_x)^2 
- b \rho \rho_{xx}  
\label{negyzelott}
=0.
\eqe 
Dividing the last equation by $b$ and multiplying by 2, we have 
\eq
 \rho_{ t_1} 
- 2 (\rho_x)^2 
- 2 \rho (\rho_{xx})  
=0
\label{negyz}
\eqe 
where $t_1=b t^2/4$. 
This equation in certain aspects may resemble to the Kardar-Parisi-Zhang(KPZ) 
equation \cite{kpz} which describes the dynamics of surface growth. 
The original KPZ equation (where the third term doesn't contain the $\rho$  variable) 
was investigated with the self-similar Ansatz by the recent 
authors \cite{barna3} and found physically reasonable solutions 
which can be expressed with the error function.

The last equation can be brought to a more compact form 
\eq 
\rho_{t_1}=(\rho^2)_{xx}, 
\eqe
There is an even more important relationship with the nonlinear heat conduction or porous media equation   
which has the form of 
\eq
\rho_{t_1} =  (\rho^m)_{xx}, \quad  m>1. 
\eqe 
  In \cite{zk} Zeldovich and Kompaneets have found the fundamental 
solution $\rho_1$ of this equation which we write in the following  form:
  \begin{eqnarray}\label{18}
    \rho_1^{m-1}=t_1^{-\alpha (m-1)}\left(A^2-B^2x^2t_1^{-2\beta}\right)_+
  \end{eqnarray}
   where $A$ is constant  and the subscript + at the bracket indicates that only physically 
  relevant - positive - solutions are taken into account. 
Regarding the other constants we have  
\begin{equation}
\alpha=\beta=\frac{1}{m+1},\quad B^2=\frac{m-1}{2m(m+1)}.
\end{equation}
 One can see that this solution has bounded support in $x$ for any $t_1>0$ 
which is a hyperbolic property. Using comparison principle for such 
equations one can show this finite speed property for 
any initial condition having compact support. However, the fronts are 
not straight lines: $x(t_1)=\pm \frac{A}{B}t_1^\beta$, 
$\beta <1$ so the speed
 of propagation $\dot{x}(t_1)$ goes to zero if $t_1$ goes to infinity. One can 
also see that $\rho_1$ is of source-type: $\rho_1(x,0)=K_1\delta(x).$  
 
 In our case $m=2$, consequently $\alpha=\beta=1/3 $, and the solution 
for the density distribution reads 
\eq
\rho(x,t_1) = t_1^{-1/3}\left(A^2-B^2x^2t_1^{-2/3}\right) = 
\left(\frac{b}{4}\right)^{-1/3} t^{-2/3}
\left(A^2-B^2x^2 \left( \frac{b}{4}\right)^{-2/3} t^{-4/3}\right). 
 \eqe
 This solution gives us a density profile for a bounded spatial region 
and after a sufficiently long time. 
The velocity field is obtained via (15) 
\eq
v(x,t) = \frac{2 x}{3 t}.  
\eqe
For the temperature distribution the 
\eq
T_t + \frac{2x}{3t}T_x = \lambda T_{xx}
\label{elozo}
\eqe
PDE have to be solved.
Considering the $ T(x,t) = t^{-\gamma}f\left(\frac{x}{t^{\omega}} \right) $
the following ODE is derived 
\eq
\lambda f'' - \frac{\eta f'}{6} +\gamma f= 0
\eqe
with arbitrary $\gamma$ and for $\omega = 1/2$. 	
The solutions are the Kummer M and  Kummer U functions of the form of
\eq
f = c_1 M\left(\frac{1}{2}-3\gamma, \frac{3}{2}, \frac{1}{12\lambda} \eta^2   \right)\eta 
  +   c_2 U\left(\frac{1}{2}-3\gamma, \frac{3}{2}, \frac{1}{12\lambda} \eta^2   \right)\eta. 
\eqe
Exhausted mathematical properties of the Kummer function can be found in \cite{abr}. 
For completeness we give the temperature distribution as well, 
\eq
T(x,t) =   t^{-\gamma} \left[ c_1 M\left(\frac{1}{2}-3\gamma, \frac{3}{2}, \frac{x^2}{12\lambda t}  \right) \frac{x}{\sqrt{t}}   +   c_2 U\left(\frac{1}{2}-3\gamma, \frac{3}{2}, \frac{x^2}{12\lambda t}   \right) \frac{x}{\sqrt{t}} \right]. 
\eqe

In contrast to the first case, we applied a simplification in the Euler equation (14) 
therefore we can consider 
solutions which are outside the scope of the self-similar class. 
Therefore, we may try to find  other physically important solutions, 
like traveling-waves with the 
Ansatz of  $\rho = h(\zeta) = h(x-ct) $  where c is usually the wave propagation velocity.   (We still use $h$ 
as shape function). 
However, if we consider (\ref{negyzelott}), 
it can be rewritten in the following form 
\eq
\rho_{t_2} -  b (\rho_x)^2 
- b \rho \rho_{xx} =0
\label{negyzutan}
\eqe
where $t_2=t^2/2$.
We take the 
\eq
\rho = h(\zeta) = h(x-ct_2) = h(x-at^2/2)
\label{wave}
 \eqe
Ansatz where instead of the constant propagation speed $c$ one may rather consider  
a kind of constant acceleration, and the constant $c$ is renotated with $a$.  
We do this in the purpose of finding analytical solutions, and to get the results 
closer to a physical interpretation. 

Inserting this formula into (\ref{negyzutan}) we obtain the following ODE: 
\eq
 b hh'' + h'(bh' +a) = 0  
\eqe
The solution is the following 
\eq
h (\zeta) = \frac{b c_1}{a}  \left [LambertW \left(   \frac{e^{\frac{-b^2 c_1 + \zeta a^2 + c_2 a}{b^2 c_1}  }}{bc_1}   \right)   +1  \right]  
\label{suruseg}
\eqe
where $c_1, c_2$ are integration constants and the Lambert-W  function, (which also called the omega function), is the inverse function of 
\eq
f(W(x)) = W(x) e^W(x)
\eqe 
and can be evaluated with the following series expansion 
\eq
W(x) = \sum_{n=1}^{\infty} \frac{(-1)^{n-1} n^{n-2}}{(n-1)!}x^n. 
\eqe
Additional properties of this function can be found in \cite{poly}.   
Banwell and Jayakumar \cite{yak} showed that a W-function describes the relation between voltage, current and resistance in a diode, and Packel and Yuen \cite{pack} applied the W-function to a ballistic projectile in the presence of air resistance. Other applications have been discovered in statistical mechanics, quantum chemistry, combinatorics, enzyme kinetics, the physiology of vision, the engineering of thin films, hydrology, and the analysis of algorithms \cite{hayes}.  Note, that in contrast to other special functions (e.g. Whitakker, Bessel, Kummer functions) the LambertW function is not so widely used or applied.   
The final form of the density field is 
\eq
\rho(x-at^2/2) =  \frac{b c_1}{a}  \left [LambertW \left(   \frac{e^{\frac{-b^2 c_1 + (x-at^2) a^2 + c_2 a}{b^2 c_1}  }}{bc_1}   \right)   +1  \right].
\label{lambwdense}  
\eqe
Figure 1 presents the density filed of (\ref{lambwdense}) for the $c_1 =c_2 = a =b =1$ parameters. Note, that the solution 
is a steep wave front. 

Inserting the space derivative of the density into  (\ref{vtransient}) and multiplying with the time the velocity field reads the following
\eq
  v(x-at^2/2)=     -\frac{ LambertW \left(   \frac{e^{\frac{-b^2 c_1 + (x-at^2) a^2 + c_2 a}{b^2 c_1}  }}{bc_1}   \right)    t } 
{  LambertW \left(   \frac{e^{\frac{-b^2 c_1 + (x-at^2)a^2 + c_2 a}{b^2 c_1}  }}{bc_1}   \right)  +1 }
\eqe
Figure 2  presents the velocity field for the $c_1 =c_2 = a =b =1$ parameters. 
Unfortunately, there is no traveling wave solution for the  temperature distribution $T(x,t) = f(\zeta) = f(x-at^2) $
because the velocity has an explicit time dependence which is transfered to the third equation of  (1) giving the following
expression
$
0  = -\lambda  f'' -  2at f' + v(\zeta)t f'  
$
which still depends on time and cannot be simplified to an ODE of $f(\zeta)$.    

\subsection{The linear EOS case}
Let's assume now the linear case of EOS where 
\eq
p = A\rho 
\eqe
where $A$ is till a constant just to fix the proper physical dimension. 
From the second case of Eq. (1)  the velocity field can be obtained for small vs 
\eq
v = -\frac{A\rho_x} {\rho}t. 
\label{kisv}
\eqe
From the continuity relation the following PDE is available for the density field 
\eq
\frac{\rho_t}{t} - A  \rho_{xx} = 0.  
\label{stand}
\eqe
Note, that the former variable transformation $t_2 = t^2/2$ is still valid giving us the regular Fourier heat conduction or 
Fick's diffusion equation of 
\eq
\rho_{t_2} - A \rho_{xx} =0. 
\eqe
The solution is the well known Gaussian function of 
\eq
\rho(x,t)  = \frac{1}{\sqrt{A t_2}}exp\left (\frac{-x^2}{2 A t_2}\right)  =  
\frac{1}{t}\sqrt{\frac{2}{A}} exp\left( \frac{-x^2}{ A t^2}\right).
\eqe
According to (\ref{kisv}) the velocity field is $v(x,t) = 2 x/t$.   
For the temperature distribution the following PDE have to be solved
\eq
T_t + 2 \frac{x}{t}T_x = \lambda T_{xx}. 
\eqe	
The solution is very similar to the former case of (\ref{elozo}). 
Using the same notation the solution reads 
\eq
f = c_1 M\left(\frac{1}{2}-\frac{1}{3}\gamma, \frac{3}{2}, \frac{3}{4\lambda} \eta^2   \right)\eta 
  +   c_2 U\left(\frac{1}{2}-\frac{1}{3}\gamma, \frac{3}{2}, \frac{3}{4\lambda} \eta^2   \right)\eta. 
\eqe
For completeness the temperature field is 
\eq
T(x,t) =  t^{-\gamma} \left[  c_1 M\left(\frac{1}{2}-\frac{1}{3}\gamma, \frac{3}{2}, \frac{3x^2}{4\lambda t}    \right)
\frac{x}{\sqrt{t}} 
  +   c_2 U\left(\frac{1}{2}-\frac{1}{3}\gamma, \frac{3}{2}, \frac{3x^2}{4\lambda t}    \right) \frac{x}{\sqrt{t}} \right].
\eqe
Figure 3 presents the Kummer M part of the whole solution with $ c_1 = \gamma = \lambda = 1$, and $ c_2 =0$ parameters. 
The Kummer U part looks very similar. 
It is straight forward that the PDE of the density field (\ref{stand}) can be investigated with the traveling wave Ansatz
(\ref{wave}), too. Resulting the next linear ODE of 
\eq
Ah'' - ah' =0. 
\eqe
The solution is obvious $h= c_1 + c_2 e^{\frac{a \zeta}{A}}$ with the density field of $\rho(x,t)= c_1 + c_2
e^{\frac{a(x-a^2t/2)}{A}}  $. It turns out from (\ref{kisv}) that $v = -at$. After trivial calculations  from the heat conduction equation we get $T(x,t) = c_1 + c_2(x-at^2/2)$. 


\subsection{Investigation of the virial EOS} 
 We still consider the original  system of (\ref{pde}), and change the  EOS in the second equation.  
Outside the ideal gas and the polytropic equation of state the most 
important material law is the Van der Waals (VdW) formula for gases  where  
$p(\rho,T) = \frac{aT\rho}{b-\rho} - c\rho^2$  
where two of the constants  ($a,b$) can be calculated from the critical point.  
One of them a is proportional with the universal gas constant  times the molar mass 
of the fluid. 
Plugging this expression back to the Euler equation and calculating the derivatives we get constrains to the exponents
 of the Ansatz $\alpha,\beta,\gamma$. For VdW EOS these constrains are contradictory, so 
no ODEs can be obtained  no self-similar solution can be evaluated. \\
We may go further and try the virial expansion 
which can be written in various forms, one possibility is the following  
\eq 
p = AT\rho(1+B\rho + C\rho^2)
\eqe
where  $A,B, C$ are constants and can be fixed from experiments. 
We try the simplest $B=C=0$ case. 
Apply the Ansatz of Eq.  (\ref{ans}).   The universality relations are therefore $ \alpha =1, \>\>  \beta = \gamma = \delta = 1/2,   \ $ which dictate 
the next ODE system
\begin{eqnarray}  
-\frac{h}{2} - \frac{1}{2} \eta h' + g h' + hg' &=& 0, \nonumber \\
-\frac{1}{2} g - \frac{1}{2} \eta g' + gg'& =& -Af'  - \frac{Afh'}{h} , \nonumber \\ 
-f - \frac{1}{2}\eta f' +gf' &=& \lambda f''.  
\label{virial}
\end{eqnarray}
Note, that the first equation is a total derivate and can be integrated  immediately which fixes $g = \eta/2$ as for the 
first investigated system in the Eq. (5).  So, $v(x,t) = \frac{x}{2t}.$  (We fixed the integration constant to zero.)
Inserting this into the second and third equation of (\ref{virial}) we get
\begin{eqnarray}  
-\frac{\eta}{4} & =& -Af' - \frac{Afh'}{h} , \nonumber \\ 
-f   &=& \lambda f''. 
\end{eqnarray}
The solution for $f$ reads: 
\eq
 f(\eta) = c_1 sin\left( \frac{\eta}{\sqrt{\lambda}} \right) +   c_2 cos\left( \frac{\eta}{\sqrt{\lambda}} \right) 
\eqe
which means 
\eq
T(x,t) = \frac{1}{t} \left[ c_1 sin\left( \frac{x}{\sqrt{\lambda t}} \right) +   
c_2 cos\left( \frac{x}{\sqrt{\lambda t}} \right)  \right]
\eqe

Finally, the formal solution for $h(\eta)$ for the density shape function can be derived  
\eq
h(\eta) = c_3 \int \cdot \frac{-A c_1 cos \left(  \frac{z}{\sqrt{\lambda}} \right)  + A c_2 sin \left(  \frac{z}{\sqrt{\lambda}} \right) +\frac{z\sqrt{\lambda}}{4}   }{ \sqrt{\lambda}A   \left[  c_1 sin \left(  \frac{z}{\sqrt{\lambda}} \right) 
 + c_2 cos \left(  \frac{z}{\sqrt{\lambda}}    \right)    \right]      } dz.  
\eqe
For given  parameters like $c_1=0,  c_2=c_3= A= \lambda=1$ the integral can be evaluated in a closed form of 
\eq
h(\eta) =  \frac{c_3(1+ie^{i\eta})^{-\frac{\eta}{4}}  (1-ie^{i\eta})^{\frac{\eta}{4}}   
  e^{-\frac{i}{4}[-4\eta -dilog(1+ie^{i\eta}) +  dilog(1-ie^{i\eta})   ]}  }{e^{2i\eta}+1}
\eqe
where $i$ is the complex unit and $dilog(x)$ is the dilogarithm (or  Spence's) function defined via the following integral
\eq
 dilog(x) =  -\int_0^x   \frac{ln|1-s|}{s} ds.
\eqe
A detailed mathematical description of the properties of the dilogarithm function can be found in \cite{abr}.
For completeness we give the final formula of the density
\eq
 \rho(x,t) = t^{-1/2} \left[  \frac{(1+ie^{i x/\sqrt{t}})^{-\frac{x/\sqrt{t}}{4}}  (1-ie^{ix/\sqrt{t}})^{\frac{x/\sqrt{t}}{4}}   
  e^{-\frac{i}{4}[-4x/\sqrt{t} -dilog(1+ie^{ix/\sqrt{t}}) +  dilog(1-ie^{ix/\sqrt{t}})   ]}  }{e^{2ix/\sqrt{t}}+1} \right]. 
\eqe
Note, that the velocity and the temperature distributions in the first and the last model are very similar to each other.  

\section{Summary and outlook} 
We investigated the basic one-dimension coupled PDE equations describing fluid flow with 
heat conduction and presented self-similar solutions for fluid density, flow
velocity and temperature.
In certain cases we have tried to find solutions for the system of equation
presented in (\ref{pde}). 
In the case when $p\sim \rho^3$ the major achievement 
is that an exact solution is available and the long time decay of 
density and pressure is a power law of the time variable. 
The other situation when $p\sim \rho^2$ is more complicated from the point of
view of exact solutions, consequently we tried to arrive to results
iteratively. In this case solutions related to waves has been also obtained 
with the help of a special function. 
The following case when $p\sim \rho$ is again a nontrivial situation 
in the frame of eq.\ (\ref{pde}).  
At this point one can arrive to solutions iteratively, which in certain aspects 
may resemble to diffusivity in dilute systems. 
Even more for long time behavior a kind of power law decay for density and pressure 
can be also observed. 

Further possible study may be the investigation of the system (\ref{pde}),  
with other possible equations of state, which may be different 
than the ones considered here. 
The extension of Eq.\ (\ref{pde}) for real fluids(e.g. including the viscous term, temperature or density dependent 
viscosity etc.) and searching for possible analytic or eventually exact solutions is also a possible future problem to be investigated. Regarding computational methods we hope that our study can help 
for benchmark tests for various numerical codes or models. 

\section{Acknowledgement} 
We thank for Prof. Robert Kersner for useful discussions.  


\begin{center}
\vspace*{0.5cm}
  \resizebox{15cm}{!}
{   \rotatebox{0} 
    {
     \includegraphics{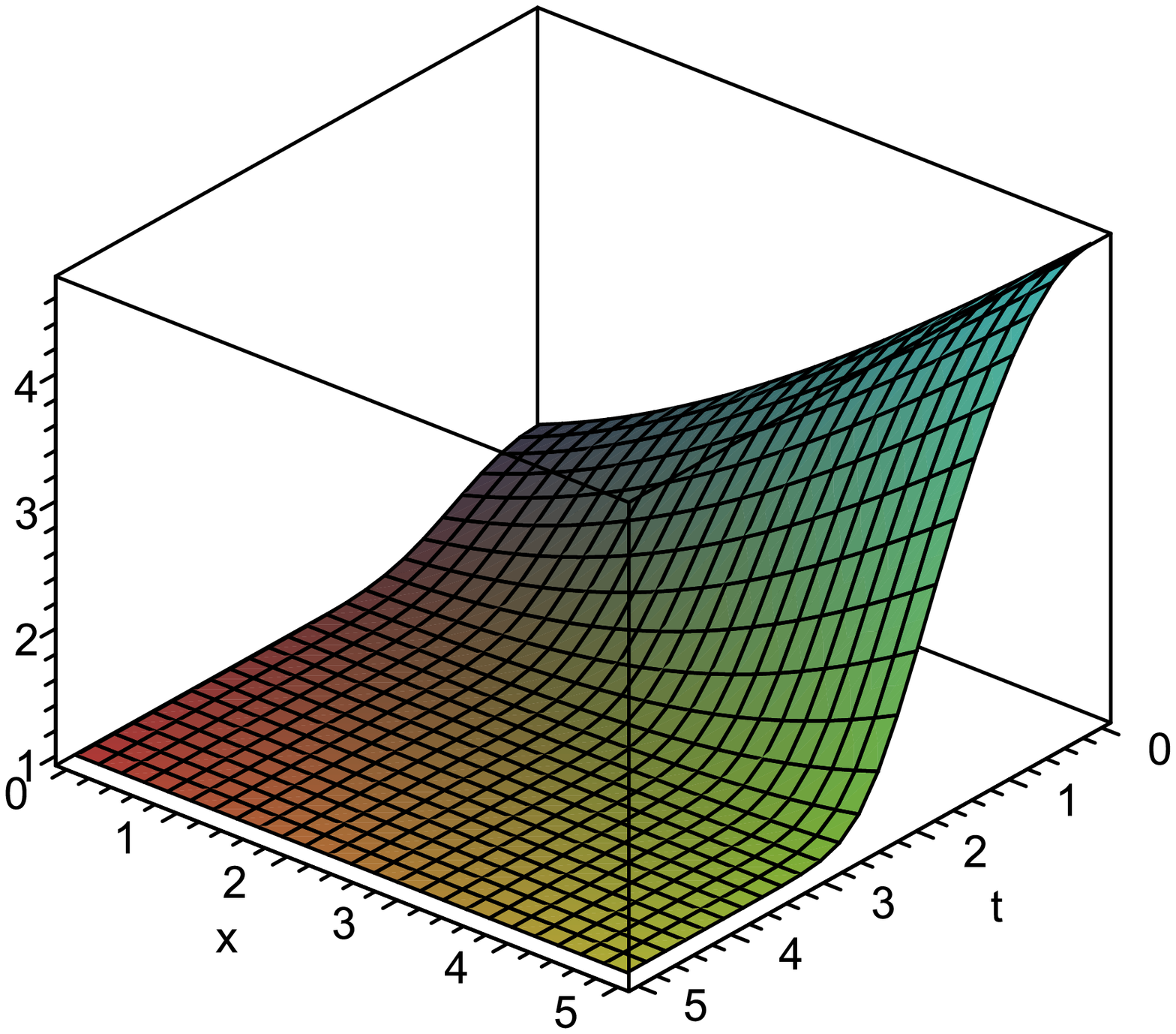}
      }}  \\
   {\bf{Fig. 1.} } The density distribution  of Eq.  35 for the  $a = b = c_1 = c_2 = 1$ parameters. 
\end{center}
\begin{center}
\vspace*{0.5cm}
  \resizebox{15cm}{!}
{   \rotatebox{0} 
    {
     \includegraphics{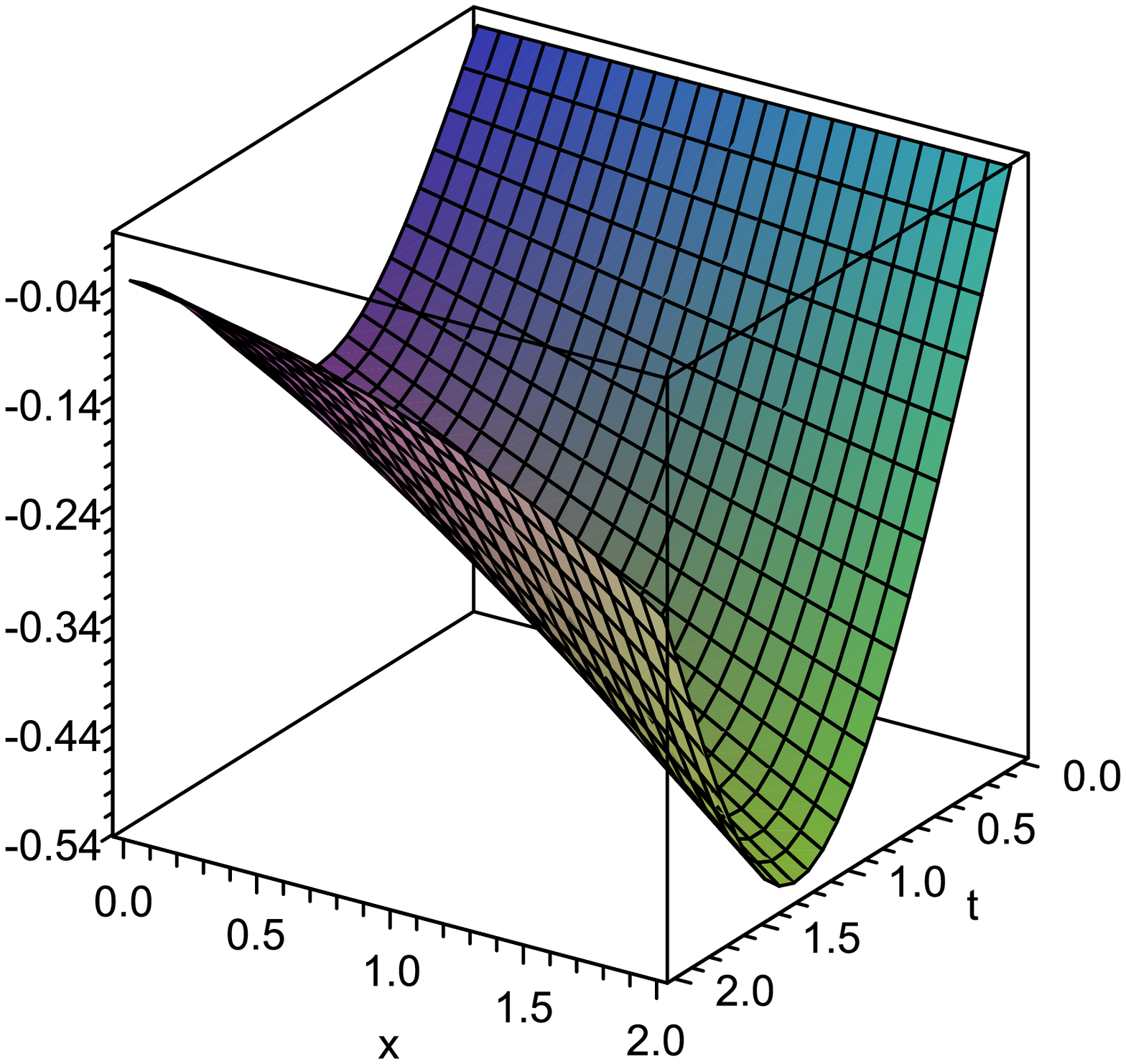}
      }}  \\
   {\bf{Fig. 2.} } The velocity field  of Eq.  36 for the  $a = b = c_1 = c_2 = 1$ parameters. 
\end{center}


\begin{center}
\vspace*{0.5cm}
  \resizebox{15cm}{!}
{   \rotatebox{0} 
    {
     \includegraphics{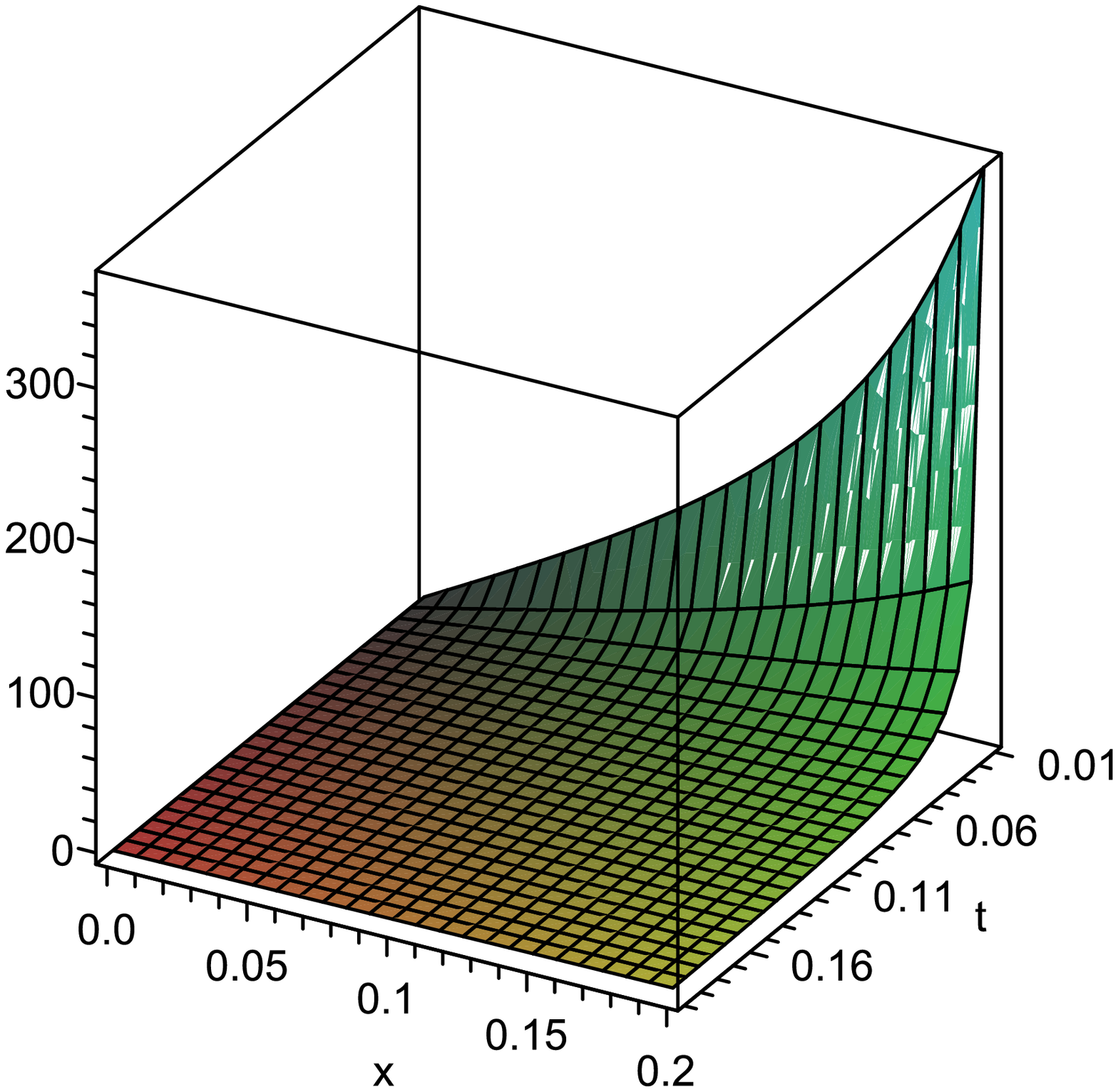}
      }}  \\
   {\bf{Fig. 3.} } The temperature field  of Eq. 44 only the Kummer M function is presented with the  $
\gamma = \lambda = c_1 =  1$ parameters. 
\end{center}



\end{document}